\documentclass[aps,pre,superscriptaddress]{revtex4}
\usepackage{amsfonts,amsmath,amssymb,latexsym,epsfig}
\usepackage{graphicx}
\begin{document}

\title{Survival Probability for the Stadium Billiard}
\author{Carl P. Dettmann \footnote{Carl.Dettmann@bristol.ac.uk}}
\affiliation{School of Mathematics, University of Bristol, United Kingdom}
\author{Orestis Georgiou \footnote{maxog@bristol.ac.uk}}
\affiliation{School of Mathematics, University of Bristol, United Kingdom}

\begin{abstract}
We consider the open stadium billiard, consisting of two semicircles joined by parallel straight sides with one hole situated somewhere on one of the sides. Due to the hyperbolic nature of the stadium billiard, the initial decay of trajectories, due to loss through the hole, appears exponential. However, some trajectories (bouncing ball orbits) persist and survive for long times and therefore form the main contribution to the survival probability function at long times. Using both numerical and analytical methods, we concur with previous studies that the long-time survival probability for a reasonably small hole drops like $Constant\times(time)^{-1}$; here we obtain an explicit expression for the $Constant$.
\end{abstract}
\maketitle

\section{Introduction}
Chaotic billiard theory was introduces by Yakov Sinai in 1970 \cite{Sin70}.
Since then it has developed to become a well established theory for dynamical systems. A billiard is a
dynamical system in which a particle alternates between motion in a straight line and specular reflections from the
domain's boundary. The sequence of reflections is described by the billiard map which completely characterizes the
motion of the particle, hence billiards have their boundaries as a natural Poincar\'{e} section. Billiard systems are convenient models for many physical phenomena, for example where one or more particles move inside a container and collide with its walls. An excellent and comprehensive mathematical introduction to chaotic billiard theory can be found in the book of Chernov and Markarian \cite{NChe}.

In the early 80's, mathematicians suggested investigating open systems, systems with holes or leakages, as a means of generating transient chaos \cite{Yorke80}, retrieving information from distributions \cite{Tamas02}, and deducing facts about the equivalent closed systems. The key distributions of interest classically are the escape probability density $p_e(t)$, which is given by the trajectories that leave the billiard at time $t$, where $t\geq0,$ and $t\in \mathbb{R}$, and also the survival probability of orbits $P(t)$ up to time $t$, given some initial probability measure, typically the equilibrium measure defined in section II below. These two are related by $P(t)=\int_{t}^\infty p_e(t')\mathrm{ d}t'$. Such investigations have naturally been extended to billiard systems as well. Links between billiards and geometrical acoustics \cite{LS90-1,LS90-2,LS90-3,LS91,LS93}, quantum chaos \cite{DS92-1,DS92-2}, controlling chaos \cite{paar97,paar00,paar01}, atom optics \cite{Milner00}, hydrodynamical flows \cite{pierrehumbert94,Picard00,schneider04,schneider04-2,schneider05,schneider07}, astronomy \cite{nagler.astro04,nagler.astro05} and cosmology \cite{motter01}, have been established in the context of open dynamical billiards. Furthermore, it has become apparent over the past few years, that the subject of open billiards and their distributions provide a pathway towards understanding chaos and may even open doors to old, but not so forgotten problems such as the Riemann hypothesis \cite{BD05,BD07}.

The stadium billiard (see Figure ~\ref{fig:stadium} below) is a seemingly simple dynamical system, introduced by Leonid Bunimovich in 1974 \cite{Bu74}. The billiard's boundary consists of two parallel straight lines and two semicircular arcs. It was later proven by him to be ergodic, to be mixing, to have the Kolmogorov property \cite{Bu79}, and in 1996 by Chernov and Haskell to have the Bernoulli property \cite{NChe96}. It has been described as a system with ``fully developed chaos" \cite{Bleher89}. Its entropy has been numerically estimated in \cite{Ben} and theoretically in \cite{Ely}. The stadium billiard is a special case of a chaotic billiard. Being constructed from two fully integrable billiard segments, the circle and the rectangle, it is remarkable that the system remains completely chaotic no matter how short its parallel segments are. It is a limiting case of the larger set of Hamiltonian systems, which Bunimovich refers to as ``mushrooms" \cite{Bu01} with cleanly divided phase-space areas, regular and chaotic. The stadium is the fully chaotic limit of the natural mushroom billiard, while the circle is the fully regular limit. If the parallel segments are of length $2a$ say, where $a>0,$ and $a\in \mathbb{R}$, then the Lyapunov exponent $\lambda(a)\rightarrow 0$ in both limiting cases of $a\rightarrow0$ and $a\rightarrow\infty$. Also, it is well known that the defocusing mechanism, which is one of the two sources of chaos in billiards \cite{Bu91} (the other being the dispersing mechanism), characteristic of all Bunimovich type billiards, requires $a>0$ in order for any wave-front to defocus and therefore exhibit hyperbolicity. Wojtkowski in 1986 \cite{Wojt86} clarified much of the mechanism behind this hyperbolic behavior.

The very existence of the parallel segments of the boundary is also the source of the intermittent behavior found in the stadium billiard. They allow for the existence of a set of marginally stable periodic orbits of zero measure but indeed of great importance. They are the main reason why the stadium is not uniformly hyperbolic. Also, though it is classically and quantum mechanically ergodic, it does not have the property of unique ergodicity \cite{Becker,Hassel08}. This means that not \textit{all} eigenfunctions are uniformly distributed and therefore this causes scarring \cite{Gabriel}. This is due to the existence of the so called ``bouncing ball" orbits, sometimes called ``sticky orbits". Semi-classically, they have caused much trouble in the treatment of the system as explained in great detail by Tanner \cite{GTan} since they affect the stability of periodic orbits close to them but do not contribute to individual eigenvalues in the spectrum of the stadium.

Lai-Sang Young's infinite Markov extension construction called a Young tower in 1998 \cite{Young98,Young99} triggered a series of rigorous mathematical proofs concerning the long time statistical properties of the stadium billiard. In 2004 Markarian \cite{Markarian04} proved that asymptotically the billiard map in the stadium has polynomial decay of correlations of order $(\log n )^{2} n^{-1}$ (here $n$ is the number of iterations of the billiard map). This method was then simplified and generalized by Chernov and Zhang in 2005 \cite{Zhang05} to include for example the drive-belt stadium where the straight segments are no longer parallel. B\'{a}lint and Gou\"{e}zel \cite{Gouzel} in 2006, used this method to prove that the Birkhoff sums of a sufficiently smooth generic observable with zero mean in the stadium, satisfy a non-standard limit theorem where its convergence to a Gaussian distribution requires a $\sqrt{n\log n}$ normalization. In 2008, Chernov and Zhang sharpen their previous estimate by removing the $\log n$ factor \cite{Zhang08} and B\'{a}lint and Melbourne show that these relations hold for observables smooth in the flow direction as well (this excludes position and velocity)\cite{Balint08}. These results for the rate of decay of correlations can, at least heuristically, be transferred into the context of the open stadium to address problems such as escape rates and survival probabilities.

Therefore, even though the stadium billiard has been observed to exhibits strong chaotic properties for short times such as approximate exponential decays of the escape times distribution and decays of correlations of initial conditions both numerically and experimentally, it has also been shown to experience a cross-over at longer times, towards an asymptotic power-law behavior \cite{chirikov,Zasl}. Hence, the stadium billiard is an example of a transient chaotic system which exhibits intermittency. Intermittency (described in more detail in section II), in open systems is a relatively new subject and is increasingly being discussed and researched in the context of non-linear Hamiltonian systems. These investigations tend to focus on the asymptotic behavior of distributions and often use the stadium as one of their main examples \cite{APik,EGA}. In fact, it has recently been suspected that the very long regular flights present in the expanded stadium are the reason why numerically, at least, the moments of displacement diverge from the Gaussian \cite{Zasl08}. It is arguably an ideal model for studying the influence of almost regular dynamics near marginally stable boundaries both theoretically \cite{GTan} and numerically \cite{Brumer91}. Armstead, Hunt and Ott \cite{Arm04} have carried out a detailed investigation into the asymptotic stadium dynamics and show that $P(t)\sim \frac{\mathrm{Const}}{t}$ for long times but do not calculate the \textit{Constant}.

In this paper we explicitly calculate the measure of the set of orbits causing the asymptotic power-law behavior, and obtain an analytic expression for the \textit{longer times} survival probability function of the stadium billiard. In contrast with Ref.\cite{BD05}, we do not assume or require that the hole is vanishingly small and in contrast to Ref.\cite{Arm04}, we do not use a probabilistic description of the dynamics. The paper is organized as follows. In section II we set the stage and state the reasoning as well as the main ideas of this paper and we also set up our problem and define all the variables and sets required. In section III and IV we consider the two main sets of initial conditions which contribute to the survival probability at long times while in section V we introduce and explain the details of the approximation method used. Finally, in section VI we present our numerical results from computer simulations and compare with the analytical ones. Conclusions and discussions appear at the end, in section VII, where future work is also discussed.

\section{The main ideas and Set-up}
\begin{figure}[h]
\begin{center}
\fbox{
\includegraphics[scale=0.4]{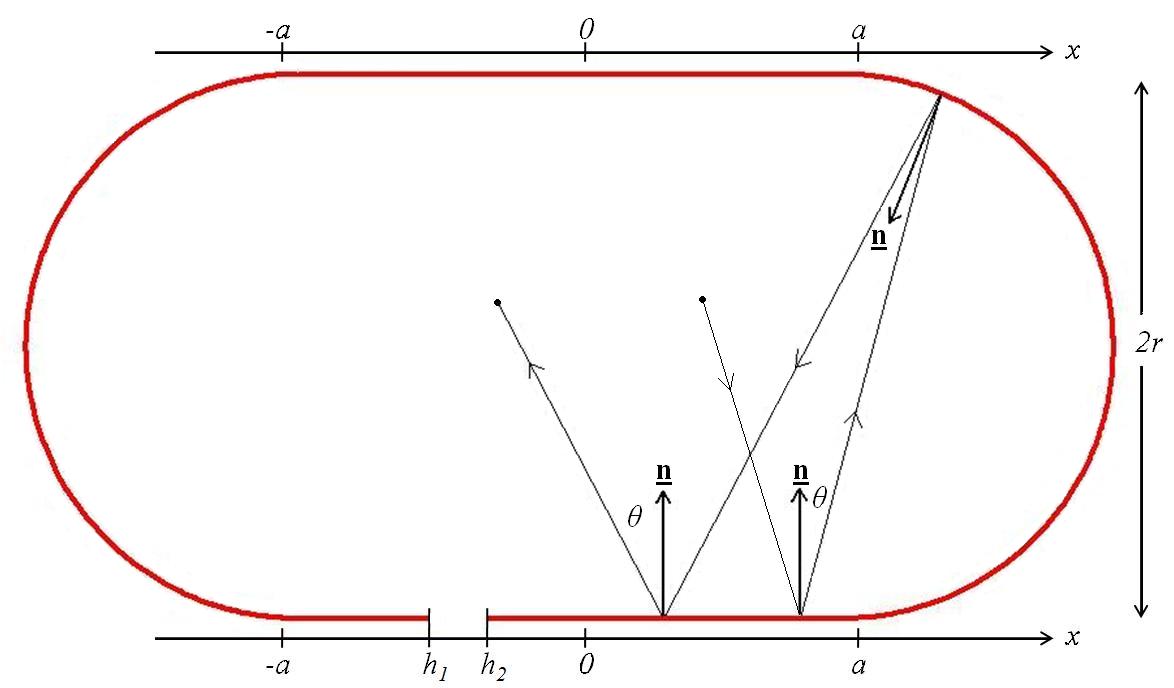}}
\caption{\label{fig:stadium}\footnotesize The Set-Up of the stadium billiard.}
\end{center}
\end{figure}

Consider an open stadium as shown in Figure ~\ref{fig:stadium}. A classical non-interacting particle of unit mass and unit speed experiences elastic collisions on $\Gamma$, the boundary of the billiard table. The length of the parallel sides is $2a$, while the radius of the circular segment is $r$. A hole of size $\epsilon$ is punched onto one of the straight segments of $\Gamma$ with $x$ coordinates $x\in(h_{1}, h_{2})$, $h_{2} = h_{1}+\epsilon$. $x$ is the position coordinate which we only need to take values along the straight segments, $x\in[-a, a]$. We define the inward pointing normal vector $\hat{\mathbf{\underline{n}}}$ which defines the angle $\theta$ made by the reflected particle and $\hat{\mathbf{\underline{n}}}$. $\theta \in (-\pi/2, \pi/2)$ , and is positive in the clockwise sense from $\hat{\mathbf{\underline{n}}}$. The billiard flow conserves the phase volume and the corresponding invariant equilibrium measure along the straight segments is $\mathrm{ d}\mu= C^{-1}\cos\theta\mathrm{ d}\theta\mathrm{ d}x$, where $C= \int_{-\pi/2}^{\pi/2}\cos\theta\mathrm{ d}\theta \int_{\Gamma}\mathrm{ d}r = 2|\Gamma| = 2\big(4a+2\pi r\big) $ is the canonical probability measure preserved by the billiard map on the billiard boundary. $\mathrm{ d}\mu$ is also the distribution of initial conditions. For the purpose of this paper, no parametrization of the position coordinate is needed along the circular segments of the billiard. Finally, if the particle hits the hole, it will escape; as noted in the Introduction, we are interested in the long time behavior of the survival probability.

As discussed in the Introduction, due to the chaotic, ergodic nature of the stadium billiard, orbits are hyperbolic almost everywhere and thus initially escape in an exponential manner through the hole as they soon come to occupy all regions of the billiard phase space. This is a result of the defocusing mechanism. There exists however a small set of parabolic, non-isolated periodic orbits \cite{Berry78} called bouncing ball orbits which is of zero measure. It has been observed that orbits in the chaotic region of the phase space which are close to these marginally stable periodic orbits show almost regular behavior. This effect is called intermittency and was first discovered in 1979 by Manneville and Pomeau in their study of the Lorenz system \cite{Pomeau79,Pomeau80}. Usually, intermittency signifies a vanishing Lyapunov exponent for unstable periodic orbits approaching regular regions in phase space \cite{Schuster}. Trajectories almost tangent to the circular arcs (``whispering gallery orbits, or rolling orbits") are unstable and of bounded path lengths and therefore do not contribute to long time tails hence introducing no intermittency effects \cite{GTan}. Near bouncing ball orbits on the contrary are not bounded, and are generally believed to be the only cause of intermittency therefore exclusively determining the asymptotic tail \cite{Vivaldi83,Arm04} of the survival probability $P(t)$ and therefore play a significant role in the stadium dynamics. Furthermore the recent work by B\'{a}lint and Melbourne \cite{Balint08} suggests that the polynomial decay of correlations in continuous time flow is due to the stadium's bouncing ball orbits and not the rolling ones. These orbits are characterized by small angles $\theta$ (near vertical) that remain small for relatively long periods of time. Chernov and Markarian describe them in their book \cite{NChe} as orbits with a large number of `\textit{nonessential collisions}'. Semi-classically, it has been suggested that an `\textit{island of stability}' surrounds this marginally stable family. Its boundary depends explicitly on $\hbar$ and the measure of this island shrinks to zero (compared with the total volume) in the semiclassical limit $\hbar\rightarrow 0$ \cite{GTan}.

Having noted, following \cite{Vivaldi83,Arm04}, that the set of orbits surviving for long times is contained in the near bouncing ball orbits, with small angles $\theta$ and position on the straight segments, we now categorize these orbits into two simple families: orbits initially moving towards the hole, and orbits initially moving away from the hole. We would like to identify the set of orbits from these two families which do not escape until a given time $t$. An important result by Lee, that dates back to 1988 \cite{Lee88}, states that the angle of a near bouncing ball orbit in the stadium remains small after a reflection with a semicircular segment. In fact, as will be explained in section IV, small angles can change by at most a factor of $3$ after being reflected off the curved billiard boundary. Therefore, given a sufficiently large time constraint $t$, surviving orbits are restricted to angles insufficient to `jump over' the hole, even after a reflection on the circular part of the billiard boundary. In this way we identify the surviving orbits as members of time dependent, monotonically shrinking subsets of the two families of orbits defined above. The measure of these subsets tends to zero as $t\rightarrow\infty$. These two subsets are considered in detail in sections III and IV below and are used to calculate, to leading order, the stadium's survival probability function for long times $P(t)$ (see equation (32) below).

\section{Case I: Moving towards the hole}
We start by considering trajectories initially on the right of the hole with $x\in(h_{2}, a)$ moving towards it. These trajectories will prove to be only a part of the survival probability function for long times. However, they are essential in order to construct a complete and accurate expression for the asymptotic limit of the full survival probability function. To ensure that such trajectories will escape when they reach the hole's vicinity, they must satisfy the following condition:
\begin{equation}
|\theta| < \arctan\Big(\frac{\epsilon}{4r}\Big),
\end{equation}
hence they will definitely not jump over the hole. The set of initial conditions $(x, \theta)$ for trajectories which will escape in exactly time $t$ satisfies:
\begin{equation}
t\sin|\theta|- \delta4r \tan|\theta|\leq x-h_{2}\leq a-h_{2},
\end{equation}
where $0<\delta<1$. For long times $\delta 4r \tan|\theta|$, which is the horizontal distance from the edge of the hole to where the particle exits the billiard, will shrink to zero ($\sim 1/t$) as the set of surviving trajectories is limited to near vertical angles. Hence we drop this nonsignificant term in what follows. Notice that we will be using physical time $t$ for our calculations but the equations are set up as if considering a map between $p\in\mathbb{N}$ collisions, with $t=\frac{2rp}{\cos\theta}$. This way, we do not need to define equations for the billiard map, which in any case would just be described by the usual reflection map. From equations (1) and (2) we can deduce that the angles must satisfy:
\begin{equation}
|\theta|< \mathrm{min}\Big\{\arctan\Big(\frac{\epsilon}{4r}\Big),\arcsin\Big(\frac{a-h_{2}}{t}\Big)+ \mathcal{O}(1/t^{2})\Big\}.
\end{equation}
The second term in (3) is the dominant one for long times. This leads to the following integral for the conserved measure of the billiard map:
\begin{equation}
I_{r}=\frac{2}{C}\int_{0}^{\arcsin\big(\frac{a-h_{2}}{t}\big)+ \mathcal{O}(1/t^{2})} \Big( \int_{h_{2}+t\sin\theta + \mathcal{O}(1/t)}^{a}\cos\theta \mathrm{ d}x\Big)\mathrm{ d}\theta,
\end{equation}
where the subscript $r$ stands for \textit{right} and $C$ was defined in section II. We are integrating over the set of initial conditions, on the right of the hole, which will not escape until time $t$. Hence we are considering escape times greater than or equal to $t$. We have also dropped the modulus sign from $\theta$ and multiplied the whole expression by a factor of $2$, due to the vertical symmetry of the problem. This simplifies to
\begin{equation}
I_{r} = \frac{\big(a-h_{2}\big)^{2}}{C t} + \mathcal{O}(1/t^{2}).
\end{equation}
This result is valid for trajectories satisfying:
\begin{center}
$\arcsin\Big(\frac{a-h_{2}}{t}\Big)+ \mathcal{O}(1/t^{2})<\arctan\Big(\frac{\epsilon}{4r}\Big)$,
\end{center}
that is
\begin{equation}
t\gtrsim\frac{8ar}{\epsilon},
\end{equation}
since the supremum of $a-h_{2}$ is $2a$. We continue with this calculation by adding the analogous contribution $I_{l}$ from the small angle trajectories starting on the \textit{left} of the hole with $x\in(-a, h_{1})$ moving towards it. This operation can easily be calculated from equation (5) by simply sending $h_{1}\mapsto -h_{2}$ and $h_{2}\mapsto -h_{1}$.
$$I_{l}= \frac{\big(a+h_{1}\big)^{2}}{C t}+ \mathcal{O}(1/t^{2}). $$
Adding the two integrals gives the measure of all initial conditions moving towards the hole that survive until time $t$:
\begin{equation}
I_{r+l} = \frac{\big(a+h_{1}\big)^{2}}{C t}+\frac{\big(a-h_{2}\big)^{2}}{C t}+ \mathcal{O}(1/t^{2}).
\end{equation}

Hence, part of the canonical Survival Probability function due to the nonessential orbits initially approaching the hole from either side, for long times satisfying condition (6) is:
\begin{equation}
P_{1}(t)= \frac{\big(a+h_{1}\big)^{2}+\big(a-h_{2}\big)^{2}}{2(4a+2\pi r)t}+ \mathcal{O}(1/t^{2}).
\end{equation}
This expression is essentially a sum of contributions from two families of bouncing ball orbits, each proportional to the square of the available length.

\section{Case II: Moving away from the hole}

Numerical simulations confirm that $P_{1}(t)$ in equation (8) is indeed not the full expression for the long time survival probability function of the open stadium billiard. The set of orbits accounted for in the previous section is only a fraction of all the marginally stable periodic orbits discussed in the introduction of this paper. In this section we will be considering orbits initially moving away from the hole, so that they experience a reflection process when they collide with the right semicircular end. We only consider the right semicircular end, as we shall later use the symmetry of the stadium to see what happens at the other one. If the initial angles are small, then the final angles (after being reflected at the wings of the stadium) will remain small and therefore survive for long times and account for the remaining set of orbits and built up the long time survival probability. In this section we will investigate and identify exactly the initial conditions which survive for long times $t$. As in the previous section, this investigation is based on the assumption that equation (1) provides an upper bound on the magnitude of angles considered, therefore ensuring that the orbits considered are indeed bouncing ball orbits that survive for long times.

Throughout this and the following section we will be using $(x_{i}, \theta_{i})$, where $x_{i}=x_{1}-2rn\theta_{i}$, as the coordinates of our initial conditions which lie on the right of the hole, $x_{i}\in(h_{2}, a)$, and move away from it. Due to the stadium's symmetry, we only need consider the case $\theta_{i}>0$. We will use $(x_{1}, \theta_{1})$ to indicate the position and angle of a trajectory right after its final collision on a flat segment, while still moving away from the hole. Therefore, the next collision of such trajectories will be on the right semi-circular segment of the billiard. This helps to distinguish between the initial conditions and their transformed final values. Notice that $\theta_{i}=|\theta_{1}|$.
\begin{figure}[h]
\begin{center}
\fbox{
\includegraphics[scale=0.3]{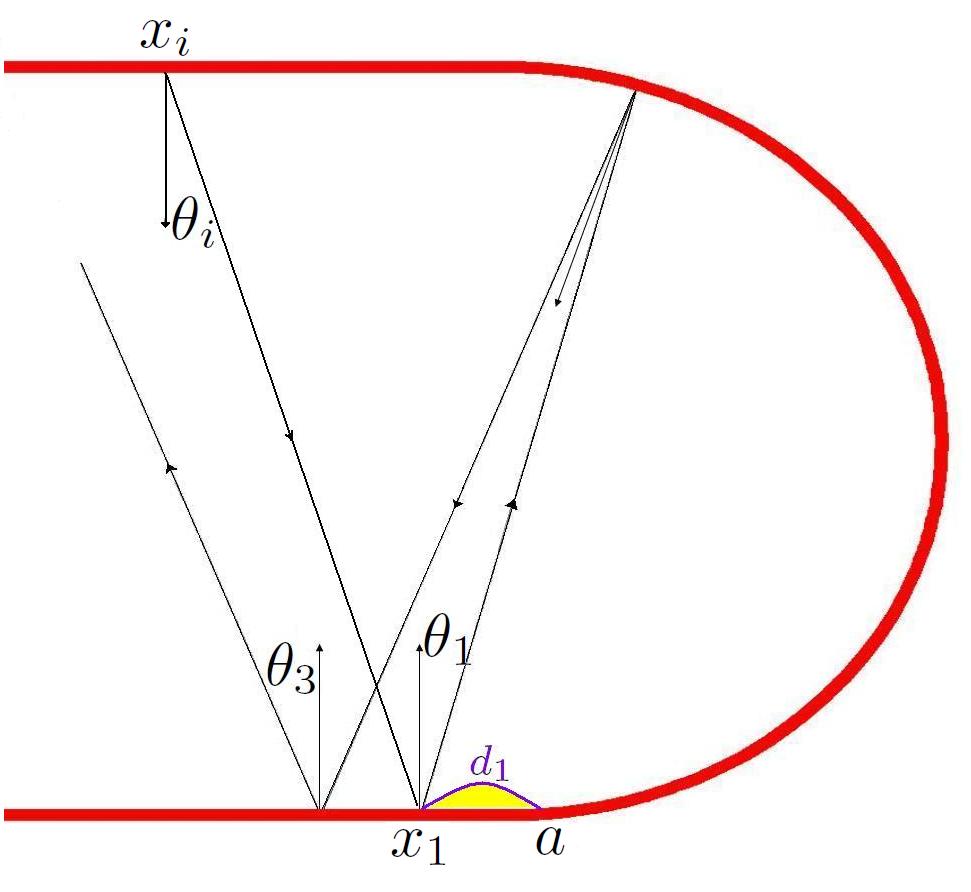}
\includegraphics[scale=0.285]{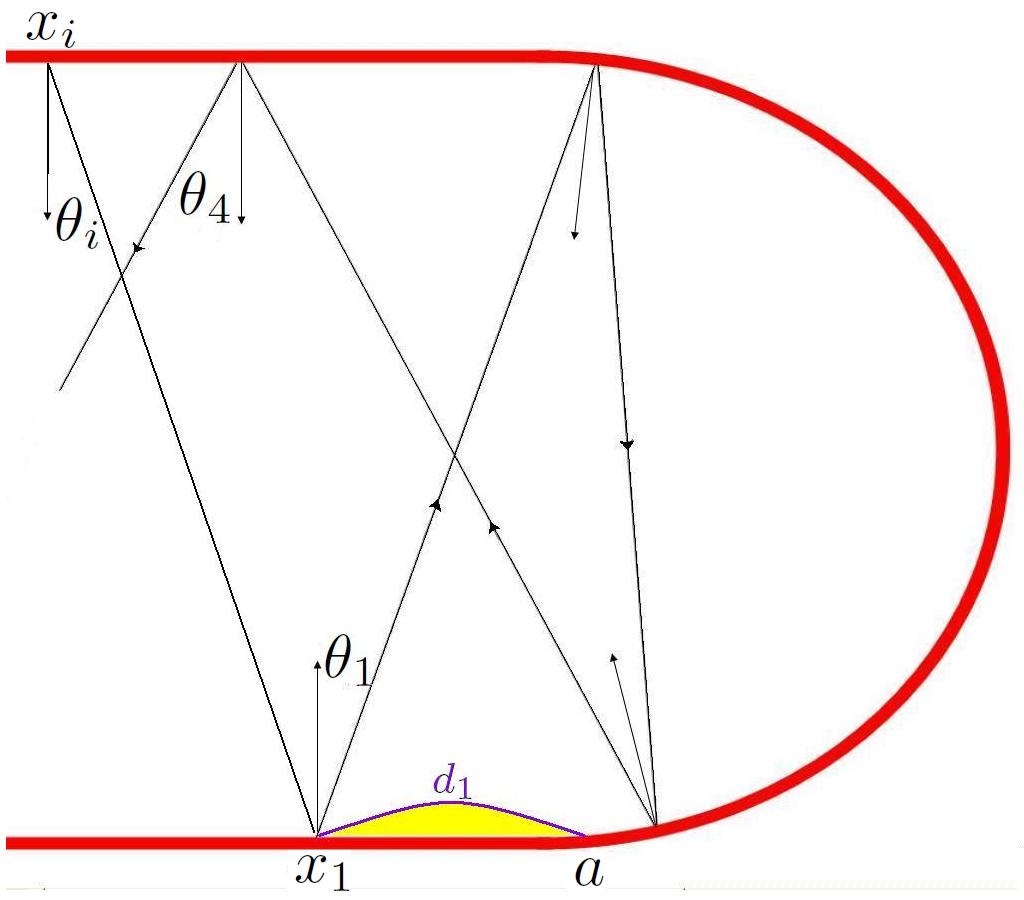}}
\caption{\label{fig:scenarios} \footnotesize The two possible scenarios of reflection from the semicircles, where $d_{1}=a-x_{1}=a-x_{i}-2rn\theta_{i} > 0$.}
\end{center}
\end{figure}

We begin by formulating the time to escape
\begin{equation}
 T(x_{i}, \theta_{i})= \frac{2rn}{\cos\theta_{i}}+\frac{2rm}{\cos\theta_{f}}+D_{f},
\end{equation}
where $n$ and $m$ are the respective numbers of non essential collisions before and after the reflection process on the right semicircular end, and are defined as:
\begin{center}
$n=\Big(\frac{a-x_{i}}{2r\tan|\theta_{i}|}-\delta_{i} \Big)$\\
$m=\Big(\frac{a-h_{2}}{2r\tan|\theta_{f}|}-\delta_{f} \Big)$,
\end{center}
with $0<\delta_{i, f}<1$, and $f= 3, 4$. We chose these indices for $f$ ($3$ and $4$) to indicate the number of collisions comprising the reflection process as the angles change two or three times respectively. $D_{f}$ is the time taken for the reflection process at the semicircular end, and is bounded by:
\begin{center}
$4r<D_{3}<2r\Big(\frac{1}{\cos\theta_{i}}+\frac{1}{\cos\theta_{3}}\Big)$\\
$6r<D_{4}<2r\Big(1+\frac{1}{\cos\theta_{i}}+\frac{1}{\cos\theta_{4}}\Big)$.
\end{center}
The number of reflections on the curved boundary alternates between two scenarios, as shown in Figure ~\ref{fig:scenarios}, the case with one collision on the semicircle and the case with two collisions depending on the initial conditions of the trajectory $x_{i}$ and $\theta_{i}$. Specifically, $\theta_{f}$ can be found and defined by the use of small angle approximations as:
\begin{equation}
\theta_{3}= \frac{2d_{1}}{r}-3\theta_{i} < 0,
\end{equation}
\begin{equation}
\theta_{4}= \frac{4d_{1}}{r}-5\theta_{i} > 0
\end{equation}
(see Ref.\cite{Arm04,Lee88}), where $d_{1}=a-x_{1}=a-x_{i}-2rn\theta_{i} > 0$, is the horizontal distance between $x= x_{1}$ and $x=a$ as indicated in Figure ~\ref{fig:scenarios}. Taking only the leading order terms of $\theta_{1}$ and $\theta_{f}$ from our expressions is justified by the fact that in long time scenarios, unstable periodic orbits have very small angles (before and after being reflected), tracing out near vertical trajectories. Using the above information, equation (9) can be expressed in the following way:
\begin{equation}
T(x_{i}, \theta_{i})= \frac{a-x_{i}}{|\theta_{i}|}+\frac{a-h_{2}}{|\theta_{f}|}+\Delta_{f},
\end{equation}
where $\Delta_{f}= D_{f}- 2r(\frac{\delta_{i}}{\cos\theta_{i}}+\frac{\delta_{f}}{\cos\theta_{f}})$.

\begin{figure}[h]
\begin{center}
\fbox{
\includegraphics[scale=0.27]{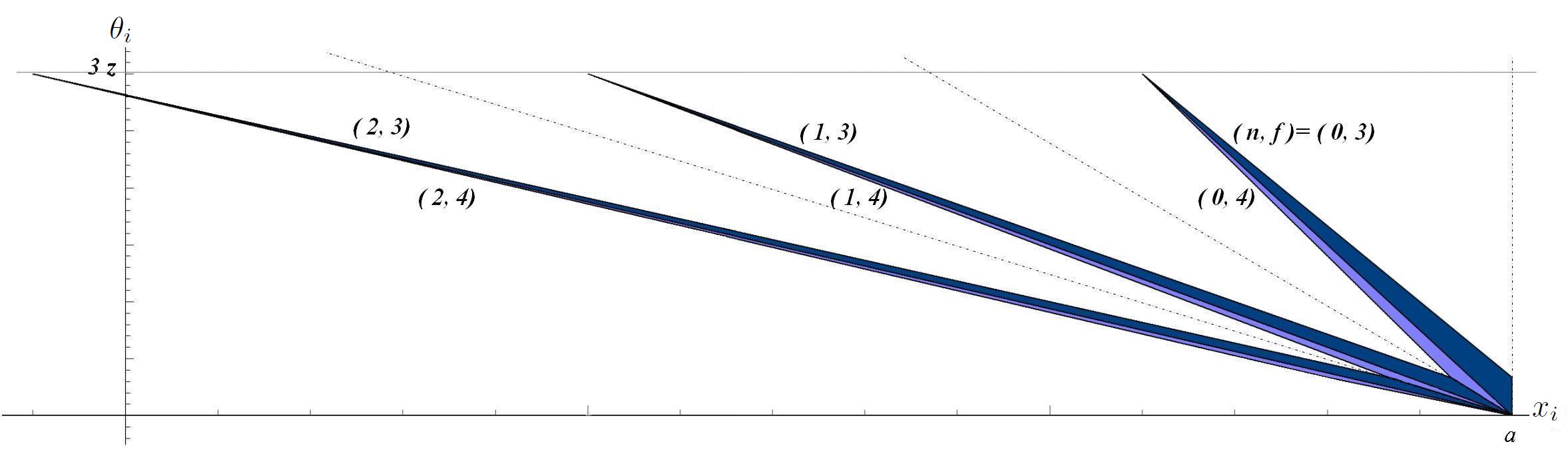}}
\caption{\label{fig:bvalidity} \footnotesize (Color online) The set of initial conditions initially on the right of the hole, which collide and reflect on the right semicircular segment of the stadium and do not cross over the hole are defined by the boundaries of validity. These are shown here for $n= 0, 1, 2$. The dotted diagonal lines are given by $\theta_{i}=(a-x_{i})/2r(n+1)$ and separate the plane into the relative areas of $n$. The top triangle of each spike (dark blue) is for $f=3$ (one collision on semicircle) while the bottom (light blue) is for $f=4$ (two collisions on semicircle). They are separated by the straight lines given by $\theta_{i}=3(a-x_{i})/2r(3n+2)$. The remaining two sets of lines which define the spikes are given by the solutions of equations (13) and (14). Notice that all spikes have the same maximum height of $3z= 3\arctan(\epsilon/4r)$.}
\end{center}
\end{figure}

Before we continue, it is essential to find the boundaries of validity for the functions of $\theta_{f}$. These will define the geometry of the two scenarios. We ask the question: When do we see one and when two collisions at the semicircular ends? This is answered by considering a number of inequalities. The first and most obvious one is $\theta_{1} \geq \frac{d_{1}}{2r}$ which requires the next collision to be on the circular segment. We also note that $\theta_{1} = \frac{3d_{1}}{4r}$ is the transition line between $\theta_{3}$ and $ \theta_{4}$ and is the case where the reflected particle will hit exactly the point $x=a$, where the circular segment meets the straight segment. Finally, if we are also to satisfy condition (1) we must form two more inequalities:\\
equation (10) gives
\begin{equation}
\theta_{i}< \frac{2d_{1}}{3r}+\frac{\arctan\Big(\frac{\epsilon}{4r}\Big)}{3},
\end{equation}
and equation (11) gives
\begin{equation}
\theta_{i}< \frac{4d_{1}}{5r}-\frac{\arctan\Big(\frac{\epsilon}{4r}\Big)}{5}.
\end{equation}
These inequalities enclose a small area in the $x_{i}\theta_{1}$ plane, the plane of initial conditions. Notice that $d_{1}$ is a function of $x_{i}$ but also depends on $n$, the number of collisions before the reflection process, which if not equal to zero introduces an extra $\theta_{1}$ term. This means that inequalities (13) and (14) have to be solved for $\theta_{1}$ for every $n=0, 1, 2, 3\ldots $ This is done and shown in Figure ~\ref{fig:bvalidity}, up to and including $n=2$, where $z$ is taken to be equal to $\arctan(\epsilon/4r)$. These boundaries of validity define the set of orbits which escape after being reflected at the right semicircular segment of the billiard. It is not immediately clear from Figure ~\ref{fig:bvalidity}, but the peaks of these spikes are of the same height $\theta_{1} = 3z$. However, we note that the set of orbits that survive up to time $T$, where $T$ is large, is not identical with the former set.

To motivate what is to follow we have a look momentarily to Figure ~\ref{fig:simulation} below, which on its left panel shows a numerical simulation which identifies the set of initial conditions $(x_{i}, \theta_{i})$ which survive until time $t = 50$. Notice that the peaks of the spikes grow in height as we move from right to left, moving away from the end of the flat segment at $x = a$, and therefore in a sense increasing the count of pre-reflection collisions $n$.
\begin{figure}[h]
\begin{center}
\fbox{
\includegraphics[scale=0.36]{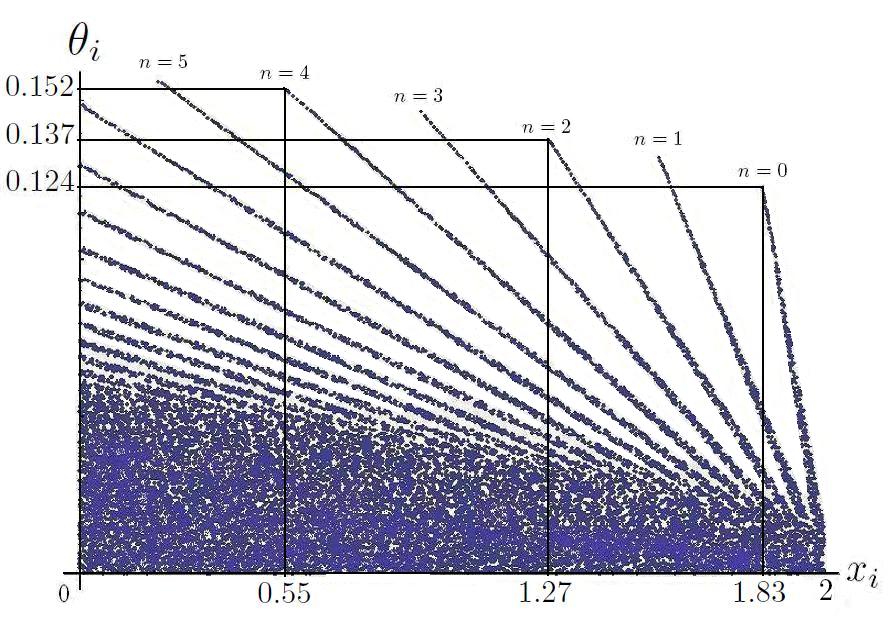}
\includegraphics[scale=0.28]{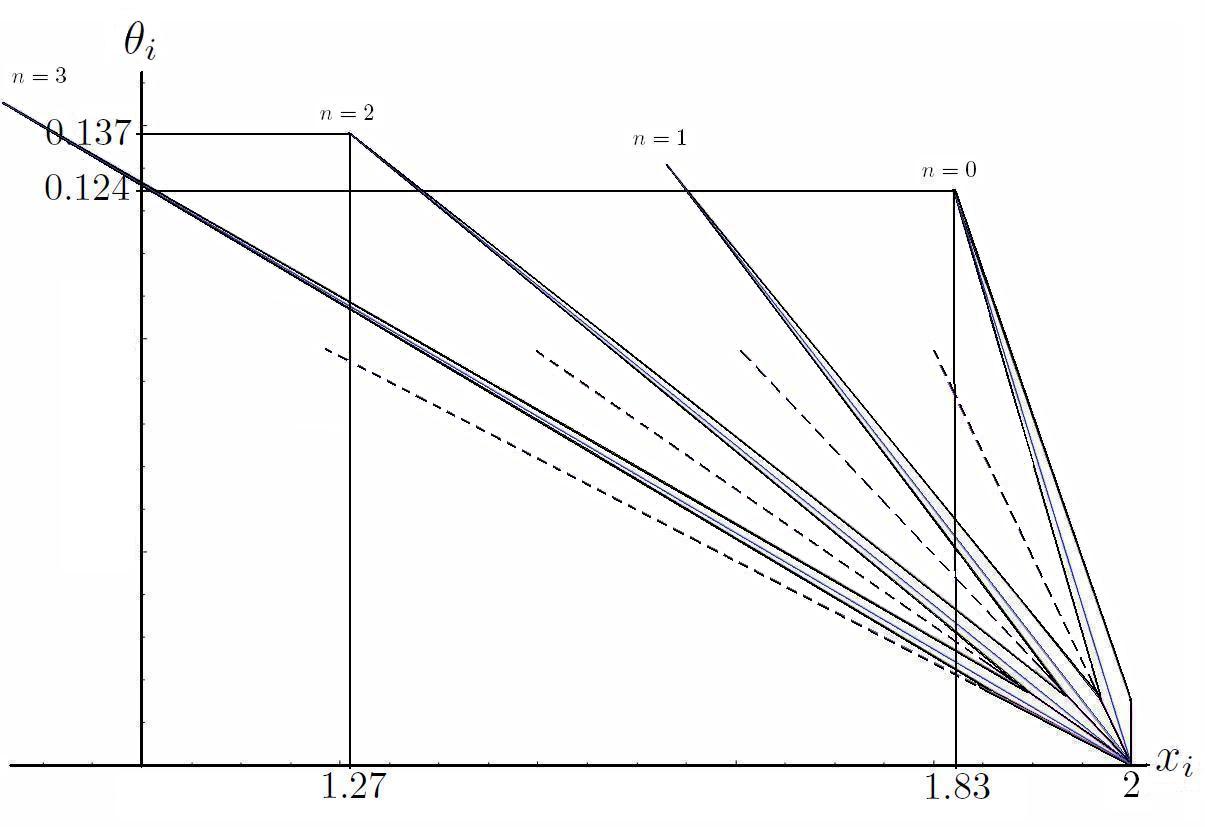}}
\caption{\label{fig:simulation} \footnotesize \emph{Left}: Numerical simulation identifying the set of initial conditions initially moving away from the hole that survive until time $t = 50$. \emph{Right}: Area enclosed by the hyperbolas for times $t =50$, for $n= 0, 1, 2$ and $3$. The line running through the middle of each spike and the diagonal lines separating them are as described in Figure ~\ref{fig:bvalidity}. Notice that unlike Figure ~\ref{fig:bvalidity}, the height of each spike is different. This is because the spikes are formed by segments of the time dependent hyperbolae defined in equations (15) and (16). This is further explained in the text. The parameters used for both left and right figures are: $a=2, r=1, \epsilon=0.2, h_{1}=-\epsilon, h_{2}=0$. The agreement of the two figures indicates that $t=50$ is sufficiently large.}
\end{center}
\end{figure}

To find the survival probability function, $P_{2}(t)$, of these trajectories we must now consider $T(x_{i}, \theta_{i})$ as the parameter $t$. Hence we can rearrange equation (12) according to (10) and (11), to form two expressions which depend not only on $x_{i}$, $\theta_{i}$ and $n$, but also on the time $t$:
\begin{align}
f_{3}(x_{i}, \theta_{i}, t)&\equiv (a-x_{i})\Big(\frac{2d_{1}}{r}-3\theta_{i}\Big)-(a-h_{2})\theta_{i}
+(\Delta_{3}-t)\Big(\frac{2d_{1}}{r}-3\theta_{i}\Big)\theta_{i}=0
\end{align}
\begin{align}
f_{4}(x_{i}, \theta_{i}, t) &\equiv (a-x_{i})\Big(\frac{4d_{1}}{r}-5\theta_{i}\Big)+(a-h_{2})\theta_{i}
+(\Delta_{4}-t)\Big(\frac{4d_{1}}{r}-5\theta_{i}\Big)\theta_{i}=0
\end{align}

The above expressions are conic sections as they are quadratic in both $x_{i}$ and $\theta_{i}$ and describe hyperbolas in the plane of initial conditions. This is not immediately obvious because of the factors $x_{i}$ and $\theta_{i}$ which are hidden in the $d_{1}$ term. The two hyperbolas, (15) and (16), approach each other as an effect of increasing the time $t$ and tilt and shift discontinuously when increasing $n$. We notice that if we impose these hyperbolas onto the boundaries of validity we found earlier (see Figure ~\ref{fig:bvalidity}), we are essentially imposing a time constraint on the set of initial conditions which will survive up to time $t$. Their effect will be for smaller $n$ to erode the area enclosed by the inequalities, therefore sharpening them, and for larger $n$ to thicken them from either side, effectively shaping them into a series of spikes, allowing for larger and larger  values of $\theta_{i}$ as we increase $n$ from left to right.

This effect can be seen on the right of Figure ~\ref{fig:simulation}, where the boundaries of validity for $n = 0, 1, 2$ and $3$ are eroded by the hyperbolas which are time dependent, causing each spike to grow taller as we move away from the edge of the straight segment of the billiard. It can be easily seen that the pictures in Figure ~\ref{fig:simulation} are almost identical to a very small error, confirming that we are measuring the correct set of initial conditions and therefore the orbits they describe. The thickening effect is a consequence of the survival probability function's set up. Trajectories which fall just outside of the area enclosed by the boundaries of validity but for large enough $n$, are trapped between the two hyperbolas, will not eventually escape through the hole, i.e. they will jump over it, but they will still survive until the given time $t$.

We notice that for finite $t$, $n$ is also finite. The key of the relation between $t$ and $n$ lies in the conic section equations (15)-(16). Solving them for $n$ we discover that for $n=\frac{-3r+2t-2\Delta_{3}}{4r}$ and $n=\frac{-5r+4t-4\Delta_{4}}{8r}$ respectively, the conic sections are no longer hyperbolas but turn into negative parabolas. Hence, we can define the maximum number of pre-reflection collisions for finite time as:
\begin{equation}
 N_{max}(t)=\Big\lfloor \rm{min}\Big\{\frac{-3r+2t-2\Delta_{3}}{4r}, \frac{-5r+4t-4\Delta_{4}}{8r} \Big\} \Big\rfloor=\Big\lfloor \frac{-5r+4t-4\Delta_{4}}{8r}\Big\rfloor
\end{equation}
for large $t$, where the lower square brackets $\lfloor h \rfloor$ are defined as the integer part of $h$ (also known as the floor function). Actually, as we shall find out in the next section, this term ($N_{max}$) is never reached in practice (see $N_{3}$ in equation (22) below).

It seems that we have the area of interest well defined and bounded. We thus need to integrate over the area of each spike and then sum them all up to $N_{max}$ for any given $t$. Multiplying the result by a factor of $2$ (vertical symmetry), would eventually give the measure of the orbits initially on the right of the hole moving away from it.

Integrating hyperbolas and then summing their enclosed areas is a lengthy and unpleasant process. This calculation has been done numerically and an accurate result has been obtained successfully and is presented in section VII. However, this calculation can only be carried out numerically, as an analytical result is in our opinion impossible to obtain. Therefore we shall present a simpler approximation method for $P_{2}(t)$, which is analytically tractable, but still accurate to leading order in $1/t$.

\section{Approximating hyperbolas}

In this section we will take equations (15) and (16) and argue that for large enough times $t$, the sections of the hyperbolas that are of interest can be simply and accurately described by straight lines. We can visually confirm this from Figure ~\ref{fig:simulation}; however, further investigations have shown that the distance between the foci of each hyperbola converges to zero faster than the lengths of the integration limits (on $\theta_{i}=(a-x_{i})/2rn$ and $\theta_{i}=(a-x_{i})/2r(n+1)$) as $t\rightarrow\infty$. In fact, for any $n$ we find that the corresponding rates are $\sim t^{-1.5}$ and $\sim t^{-1}$. This in turn shows that the error made by approximating hyperbolas by straight lines, after integrating and summing over $n$, is still negligible with respect to the leading term ($t^{-2}$ rather than $t^{-1}$). This approximation method will later be verified by further numerical simulations where we calculate the error between the approximate solution and the numerical integration result.

\begin{figure}[h]
\begin{center}
\fbox{
\includegraphics[scale=0.38]{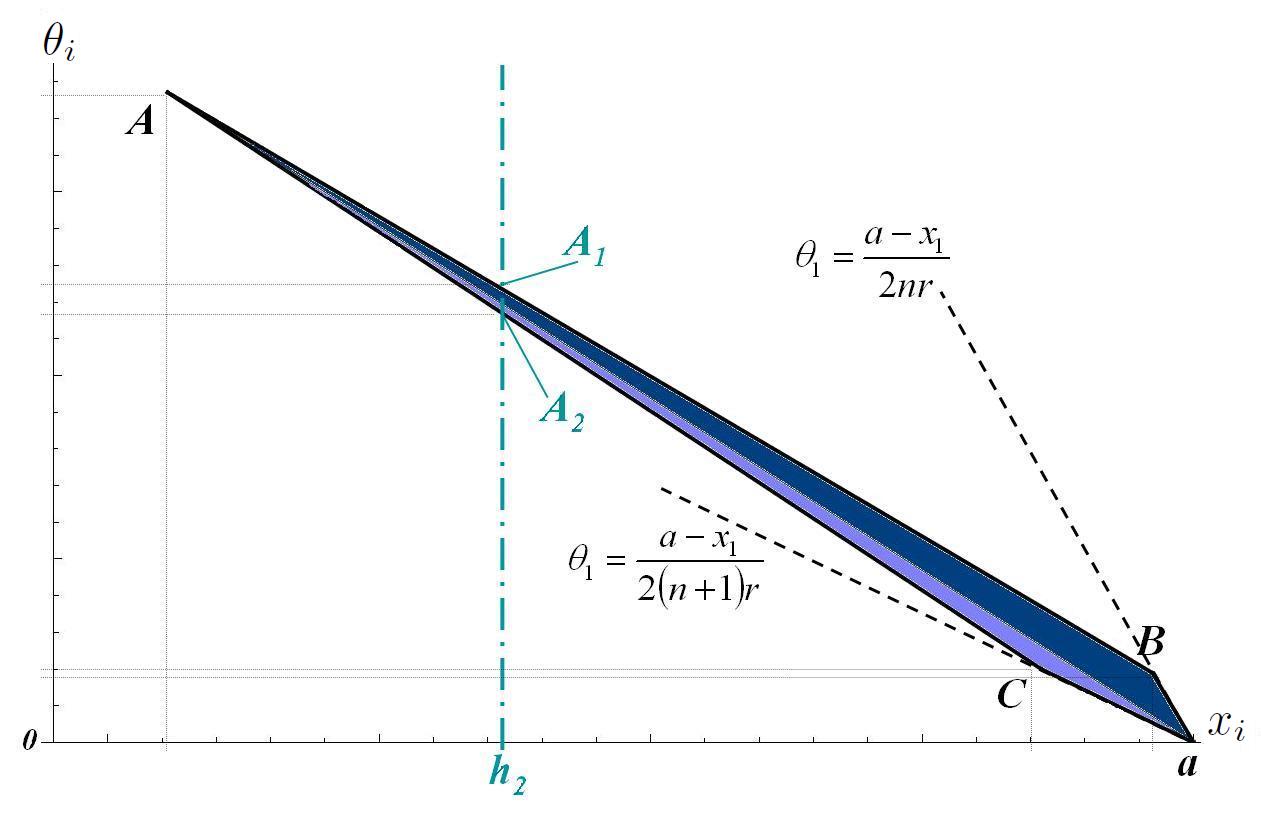}}
\caption{\label{fig:coordinates} \footnotesize (Color online) The corners which make up the polygons which approximate the hyperbolas, for any $n$, are defined by the coordinates of points $A, B, C$ and the corresponding coordinate of the endpoint of the straight segment (in this case $(a,0)$). The $x_{i}$ and $\theta_{i}$ coordinates of these corners are clearly marked by the dotted lines. The color coding is similar to that of Figure ~\ref{fig:bvalidity}. The green/dashed vertical line indicates the hypothetic position of the closest edge of the hole (in this case $x_{i} =h_{2}$) which acts and changes the shape of the spike by defining two new corners $A_{1}$ and $A_{2}$ instead of $A$.}
\end{center}
\end{figure}

In this section we will use $\tau = t-\Delta$ without any subscript to avoid unnecessary confusion given that for long times, $\Delta$ will completely vanish from our results. We consider the same initial conditions as in the previous section. For the first step of the approximation method, we must find the coordinates where equations (15) and (16) meet with $\theta_{i} = \frac{3(a-x_{i})}{2(1+n)r}$. We also need to find the coordinates where equation (16) meets with $\theta_{i} = \frac{a-x_{i}}{2(1+n)r}$ and finally equation (15) with $\theta_{i} = \frac{a-x_{i}}{2nr}$. These three points along with $(a, 0)$ define the four corners of a quadrilateral in phase space shown in Figure ~\ref{fig:coordinates}. Here are their coordinates:
\begin{align*}
A &= \Big(a+\frac{6(a-h_{2})(2+3n)r}{(4+6n)r-3\tau}, \frac{9(a-h_{2})}{3\tau-(4+6n)r} \Big),\\
B &= \Big(a+\frac{2(a-h_{2})nr}{3(2nr-\tau)}, \frac{(a-h_{2})}{3(\tau-2nr)} \Big),\\
C &= \Big(a+\frac{2(a-h_{2})(1+n)r}{3(2(1+n)r-\tau)}, \frac{(a-h_{2})}{3(\tau-2(1+n)r)} \Big).
\end{align*}
The next step is to form four equations, one for each side of the quadrilateral by using these coordinates for $n \geq 0$. Afterwards, elementary integration methods in $x_{i}$ are used to produce explicit functions for the area of each spike with $\tau$ and $n$ as the only parameters:
\begin{equation}
 Area_{1}=\frac{(a-h_{2})^{2}r}{(2nr-\tau)\big(2(1+n)r-\tau)\big)}.
\end{equation}

Before we insert equation (18) into a sum, we must figure out the upper limit of $n$ for which this expression is valid. We do this by finding the smallest integer value of $n$ for which the $x_{i}$ coordinate of point $A$ is smaller than $h_{2}$ and call it $N_{1}$. This is because, for increasing values of $n$, all the corners ($A, B, C$) shift to the left causing the spike to tilt and stretch but when $n \geq N_{1}$ point $A$ is no longer a valid coordinate. A closer look at the situation reveals that point $A$ will split in to two points, $A_{1}$ and $A_{2}$ say, both situated on the line $x_{i}=h_{2}$. For $n>N_{1}$, the quadrilateral is replaced by a pentagon as the spike's peak (point $A$) overshot the vicinity of the hole's location. This process is best described diagrammatically in Figure ~\ref{fig:coordinates} where the four corners $A, B, C$ and $(a,0)$ are replaced by $A_{1}, A_{2}, B, C$ and $(a,0)$ which are then fed back to the integration method to produce a new expression describing the area of the truncated spike. The same will happen to all corners, and different combinations of them will be necessary to produce the corresponding area functions. In other words, as the $x_{i}$ coordinates of corners $A, C$ and $B$, in this order, overshoots the hole's location at $x_{i}=h_{2}$, as we increase $n$, a new area function ($Area_{j}$, $j= 1 - 4$) via integration, with a new expiration number (summation limit) via equation solving in $n$, will be required. This process produces three more area functions, and therefore four summations, each with different limits:
\begin{align}
&Area_{2}=
\nonumber \\
&-\Bigg((a-h_{2})^{2}\Big[1152n^{4}r^{4} + (-224r^{3}+313r^{2}\tau -114 r \tau^{2}+9\tau^{3})\tau+192n^{3}r^{3}(14r-9\tau)
\nonumber \\
&+4n^{2}r^{2}(512r^{2}+225\tau^{2}-762r\tau)+4nr(128r^{3}+270r\tau^{2}-45\tau^{3}-400r^{2}\tau)\Big]\Bigg)
\nonumber \\
&\div \Bigg(r\Big(16(1+n)(2+3n)r-3(5+8n)\tau\Big)(2nr-\tau)\Big(2(1+n)r-\tau\Big)\Big[-9\tau+4n\Big((4+6n)r-3\tau\Big)\Big]\Bigg),
\end{align}
\begin{equation}
Area_{3}=\frac{(a-h_{2})^{2}\Big(32nr^{2}+16n^{2}r^{2}-(14r+3\tau)\tau\Big)}{4r(1+n)(\tau-2nr)\Big[-9\tau+4n\Big((4+6n)r-3\tau\Big)\Big]},
\end{equation}
\begin{equation}
Area_{4}=\frac{(a-h_{2})^{2}}{4(n+n^{2})r}.
\end{equation}

Having all the \textit{puzzle} pieces at hand, we form an expression for the invariant measure of all the initial conditions moving away from the hole from the right. The sum over the areas of quadrilaterals ($Area_{1}$) added to the sum of pentagons ($Area_{2}$) and another sum of quadrilaterals ($Area_{3}$) and finally the infinite sum of triangles, $Area_{4}$, gives:
\begin{align}
Area_{Right} = \sum_{n=0}^{N_{1}}Area_{1} + \sum_{n=N_{1}}^{N_{2}}Area_{2}+ \sum_{n=N_{2}}^{N_{3}}Area_{3} + \sum_{n=N_{3}}^{\infty}Area_{4},
\end{align}
where
$$N_{1}= \Big\lfloor \frac{3t-3\Delta-16r}{24r} \Big\rfloor,\qquad N_{2}= \Big\lfloor \frac{3t-3\Delta-8r}{8r} \Big\rfloor,\qquad \rm{and}\quad N_{3}=\Big\lfloor\frac{3t-3\Delta}{8r} \Big\rfloor.$$
All these sums, except the second one, where simplified as follows by \textit{Mathematica v.6}, by allowing the summation limits to acquire their non-integer values. This is allowed since the upper limit $N_{l}\sim t$ ($l=1, 2, 3$), and therefore losing or gaining a term from the end of each summation effectively makes no difference whatsoever for long times.
\begin{equation}
\sum_{n=0}^{N_{1}}Area_{1} = \frac{(a-h_{2})^{2}(8r+3\tau)}{2\tau(9\tau-8r)},
\end{equation}
\begin{align}
\sum_{n=N_{2}}^{N_{3}}Area_{3} &= \frac{32(a-h_{2})^{2}r\big(192r^{2}+56r\tau+3\tau^{2}\big)}{(8r+3\tau)(-8r-\tau)\big(64r^{2}-9\tau^{2}-72r\tau\big)}, \end{align}
\begin{equation}
\sum_{n=N_{3}}^{\infty}Area_{4} = \frac{(a-h_{2})^{2}}{4rN_{3}}.
\end{equation}
The simplification of the second sum (that of $Area_{2}$) requires a more lengthy and tricky process, as it can not be simplified explicitly by any conventional means. This is so, not only because $Area_{2}$ has the most complicated of the four expressions, but also because its sum covers the largest range over $n$ ($N_{2}-N_{1}\sim t/4$). Therefore, for large $t$, $n$ is never small. By using a substitution of the form $u = 1/t$, assuming $u$ to be small for large $t$ and then substituting $n=s/u$, where $s$ is of $\mathcal{O}(1)$, before expanding $Area_{2}$ into a power series effectively incorporates the effect of large $n$ into the leading order term of the series. We get:
\begin{equation}
Area_{2} = \sum_{k = 0}^{\infty}\alpha_{k}u^{k} =-\frac{\Big((a-h_{2})^{2}(1-16sr+32s^{2}r^{2})\Big)u^{2}}{32\Big(s^{2}r(-1+2sr)^{2}\Big)}+\mathcal{O}(u^{3}),
\end{equation}
We reverse the substitution, and simplify the sum to obtain:
\begin{align}
\sum_{n = N_{1}}^{N_{2}} Area_{2} &= (a-h_{2})^{2}
\Bigg(\frac{12ru\big(\Psi^{(0)}(z_{1})-\Psi^{(0)}(z_{2})+\Psi^{(0)}(z_{3})-\Psi^{(0)}(z_{4})\big)}{32r}
\nonumber \\
&+ \frac{-\Psi^{(1)}(z_{1})-\Psi^{(1)}(z_{2})+\Psi^{(1)}(z_{3})+\Psi^{(1)}(z_{4})}{32r}\Bigg).
\end{align}
where
\begin{align*}
z_{1} &= \frac{8r- 9/u -3\Delta}{24r},\\
z_{2} &= \frac{8r+ 3/u -3\Delta}{24r},\\
z_{3} &= \frac{3(1/u -\Delta)}{8r},\\
z_{4} &= -\frac{1/u +3\Delta}{8r},
\end{align*}
and $\Psi^{(k)}$'s are polygamma functions. The polygamma function of order $k$ is defined as the $(k + 1)$th derivative of the logarithm of the gamma function:
\begin{center}
$\Psi^{(k)}(z) = \frac{\mathrm{d}^{(k+1)}}{\mathrm{d}z^{(k+1)}} \ln\Gamma(z)$.
\end{center}
Fortunately the polygamma functions are of the form $z = \frac{a}{bu} + c$, where $a, b$ and $c$ are constants, and can be expanded as a Taylor series to leading order as follows:
\begin{align*}
\Psi^{(0)}(\frac{a}{b u} + c) &= \ln(a/b)-\ln u + \mathcal{O}(u),\\
\Psi^{(k \geq 2)}(\frac{a}{b u} + c) &= (-1)^{(k-1)}(k-1)!\Big(\frac{b u}{a}\Big)^{k} + \mathcal{O}(u^{k+1}).
\end{align*}
Substituting these expressions into equation (27) will simplify the expression dramatically, finally leaving us with the desired result. We substitute $t = 1/u$ back in to get:
\begin{equation}
\sum_{n=N_{1}}^{N_{2}}Area_{2}= \frac{(a-h_{2})^{2}(9\ln3-4)}{12t} + \mathcal{O}(1/t^{2}).
\end{equation}

In light of equations (23),(24),(25) and (28), we can now simplify (22) to first order to get:
\begin{equation}
Area_{Right} =\frac{(a-h_{2})^{2}(3\ln3+2)}{4t} + \mathcal{O}(1/t^{2}).
\end{equation}
To find an expression for $Area_{Left}$ we must use the same approach used in section III to calculate $I_{l}$. This gives:
\begin{center}
$Area_{Total} = 2(Area_{Right}+Area_{Left})$,
\end{center}
\begin{equation}
Area_{Total} = \frac{(3\ln3+2)\Big( (a+h_{1})^{2} + (a-h_{2})^{2} \Big)}{2t}+\mathcal{O}(1/t^{2}).
\end{equation}
Dividing by $C$, the total measure of the billiard map, we can obtain an approximate result for the long time survival probability of all initial conditions initially moving away from the hole:
\begin{equation}
P_{2}(t) =\frac{(3\ln3+2)\Big( (a+h_{1})^{2} + (a-h_{2})^{2} \Big)}{4(4a+2\pi r)t}+\mathcal{O}(1/t^{2}).
\end{equation}

\section{Result and Numerical Simulation}

It remains to add the probability measure of the two types of trajectories to obtain the asymptotic limit of the survival probability function:
\begin{center}
$P_{s}(t)=P_{1}(t)+P_{2}(t)$,
\end{center}
where the subscript \textit{s} stands for the \textit{straight} lines we have approximated the hyperbolas with. This gives:
\begin{equation}
P_{s}(t)= \frac{(3\ln3+4)\Big( (a+h_{1})^{2} + (a-h_{2})^{2} \Big)}{4(4a+2\pi r)t}+\mathcal{O}(1/t^{2}),
\end{equation}
which is valid only for trajectories satisfying (6), i.e. sufficiently large $t$.

In the left panel of Figure ~\ref{fig:result} below, we compare $P_{s}$ (equation (32)) with $P_{d}$ which is obtained by a \textit{direct} numerical simulation using \textit{Mathematica v6.}, consisting of $1.5$ million initial conditions distributed according to the invariant measure of the billiard map. We see that $P_{s}$ gives a good prediction of the numerical survival probability for long times $P_{d}$. We have tested this result with other values of the parameters: $a, r, h_{1}, h_{2}$ as well. What is even more important however is that equation (32) is found to be an asymptotic formula. This is shown in the right panel of Figure ~\ref{fig:result}, where we have plotted the $(P_{s} - P_{h})$ at regular intervals of time, and fitted it to an inverse time curve $D/t^{2}$, where $D$ is some constant. Here, $P_{h}$ is the result obtained by numerically summing over the areas of each spike (see Figure ~\ref{fig:simulation}), found by the integrated difference of  $\theta_{3}(x_{i}, t, n)$ and $\theta_{4}(x_{i}, t, n)$ which define the \textit{hyperbolas} in the $x_{i}\theta_{i}$ plane. We find that the $(P_{s} - P_{h})$ fits perfectly into $D/t^{2}$ , where $D$ needs to be calculated by a numerical fit. Thus we confirm that the approximation chosen in section V was justifiable from the asymptotic convergence to the integral $P_{h}$. $D$ is simply the coefficient of the second order term in:
\begin{equation}
P(t)= \frac{(3\ln3+4)\Big( (a+h_{1})^{2} + (a-h_{2})^{2} \Big)}{4(4a+2\pi r)t}+ \frac{D}{t^{2}}+ o(1/t^{2}).
\end{equation}
\begin{figure}[h]
\begin{center}
\fbox{
\includegraphics[scale=0.275]{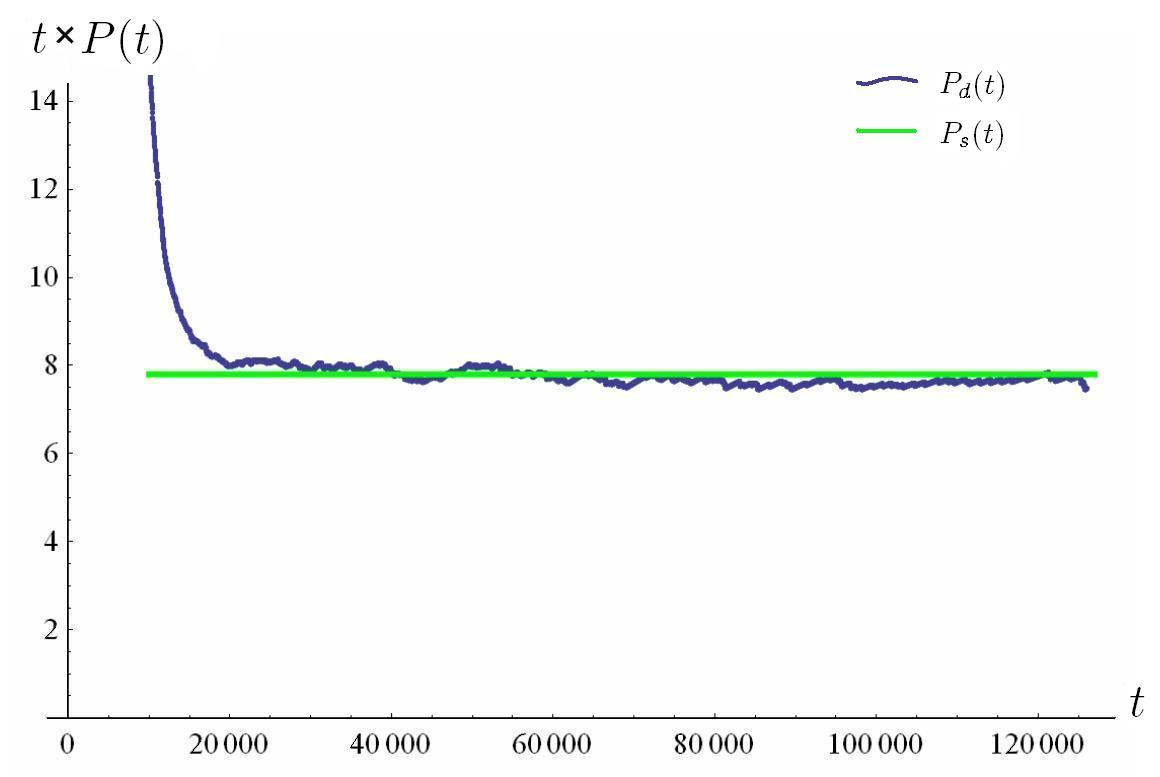}
\includegraphics[scale=0.3]{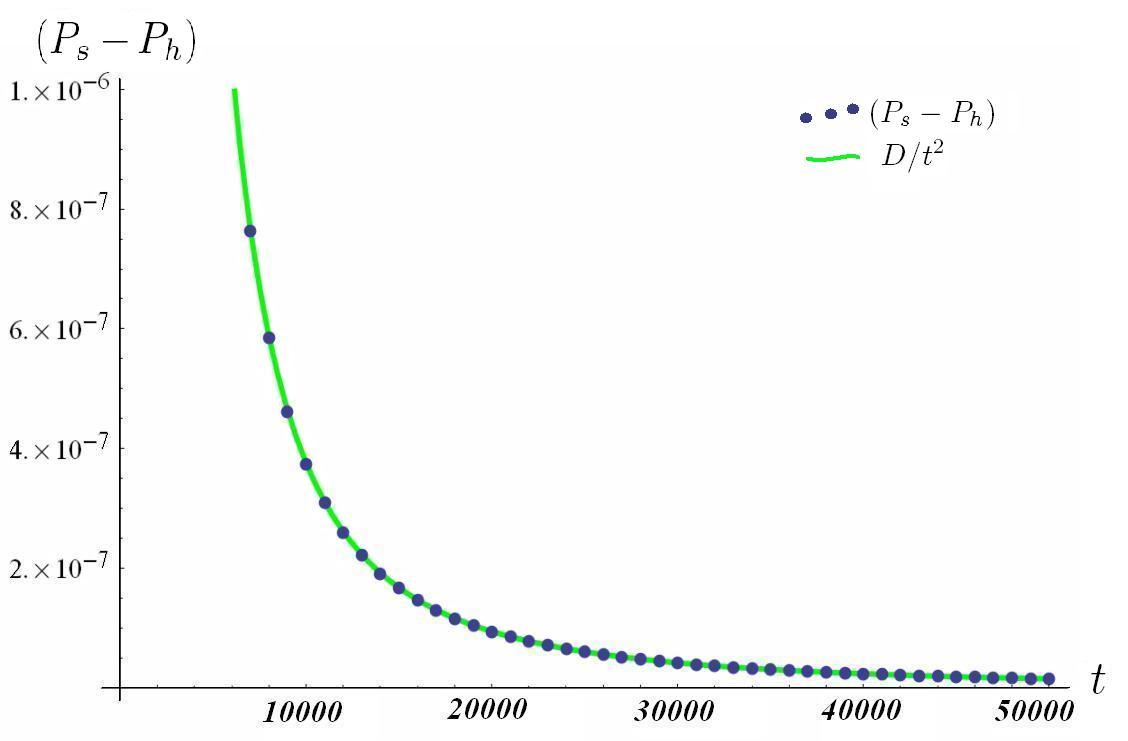}}
\caption{\label{fig:result} \footnotesize (Colour online)\emph{Left:} Plot comparing the survival probability $P_{s}$ (light green/horizontal line) found by equation (32), with the numerical survival probability $P_{d}$ (dark blue/curve), found by direct numerical simulation, both multiplied by the time $t$. \emph{Right:} The difference between equation $P_{s}$ and $P_{h}$ which is found by numerically integrating hyperbolas and summing the relative areas under the spikes created, decays as $D/t^{2}$. $D$ is the coefficient of the second order term in equation (33).}
\end{center}
\end{figure}

\section{Conclusion and Discussion}

In this paper, we have investigated the open stadium billiard and managed to derive, using phase space methods, an expression consisting of the two main contributions (see equations (8) and (31)) to the long time asymptotic tail of the survival probability function of the stadium billiard. Both expressions are to leading order in $t$. The second one (equation (31)) is an approximate result which converges to the true result as $\sim t^{-2}$ which means the errors are $\mathcal{O}(1/t^{2})$ and hence do not appear in the simple closed form of equation (32). The expression has been confirmed through numerical simulation (see Figure ~\ref{fig:result} \textit{Left}). In total, we confirm that the survival probability of the stadium for long times goes as $\frac{Constant}{t}$, and we find that the $Constant$ depends quadratically on the lengths of the parallel segments of the billiard on either side of the hole and hence the size of the hole as well as its position on one of the straight segments of the boundary (see equation (32)).

In the context of stadia, there is a variety of possible shapes for which one can observe similar properties. In this paper we only considered the standard stadium that is a construction of two parallel straight lines and two complete semicircular arcs. It is also possible to construct different ergodic stadia by using circular arcs of lengths less than $\pi r$ or by using elliptical arcs \cite{Markarian95}. In both cases we expect ergodicity and an initial strong decay of the survival probability followed by an asymptotic power law decay at longer times, provided that the parallel straight sides are still present, and that the dynamics remain defocusing \cite{Bu08}. Hence, similar methods used in the present paper should be applicable to some variations of the stadium geometry as explained above.

At this point, we would like to comment on the $\ln 3 $ term, which first appeared in equation (28). Similar terms where found in the work of both B\'{a}lint and Gou\"{e}zel \cite{Gouzel} and Armstead's \textit{et al.} \cite{Arm04} as well as several other papers relating to the stadium's bouncing ball orbits and its long time dynamics. It appears, that the  $\ln 3 $ term is a direct consequence of the geometry of our stadium billiard. More specifically, the circular curvature of the boundary near the straight segments, leads to a reflected final angle $|\theta_{f}|\in(|\theta_{i}|/3,3|\theta_{i}|)$, if $\theta_{i}$ is small enough; this follows from equations (10) and (11) above. Hence we propose that any bounded change in the curvature of the focusing segments of the billiard, such that the boundary remains $C^{1}$ smooth, would change the dynamics quantitatively but not qualitatively (\textit{i.e.} $P(t)\sim\frac{Const}{t}$).

Further work on this subject may include researching the open stadium with holes on the circular segments. Such an example is expected to behave very similarly to the case described in the present paper as is numerically shown in \cite{Brumer91}. This is because the trajectories which dominate and survive for long times, again will be characterised by small (near vertical) angles. Their collisions will mainly be with the straight segments of the billiard, but also on very short segments of the semicircular arcs. What is obviously different in such a situation is that the number of collisions with the semicircular arcs is not restricted to only one, as was the case here. This fact will complicate the dynamics substantially. Therefore one might prefer to choose a probabilistic approach to such a problem, as suggested by in Armstead's \textit{et al.} \cite{Arm04}, rather than an analytic one. One might also be interested in addressing this from a different perspective such as a semiclassical approach or even a quantum mechanical one and hopefully obtain some sort of correspondence between results.

\section*{Acknowledgements}
We would like to thank Uzy Smilansky for helpful discussions, and OG's EPSRC Doctoral Training Account number SB1715.

\end{document}